\documentclass[a4paper]{quantumarticle}
\pdfoutput=1
\usepackage[utf8]{inputenc}
\usepackage{amsmath,amssymb,amsthm}
\usepackage{graphicx}
\usepackage{booktabs}
\usepackage{hyperref}

\graphicspath{{./}}

\title{Fair Decoder Baselines and Rigorous Finite-Size Scaling for Bivariate Bicycle Codes on the Quantum Erasure Channel}

\author{Tushar Pandey}
\orcid{0000-0001-7448-5723}
\affiliation{Department of Electrical \& Computer Engineering, Texas A\&M University, College Station, TX 77843, USA}
\email{tusharp@tamu.edu}

\date{}

\begin{document}

\maketitle

\begin{abstract}
Fair threshold estimation for bivariate bicycle (BB) codes on the quantum erasure channel
runs into two recurring problems: decoder-baseline unfairness and the conflation of
finite-size pseudo-thresholds with true asymptotic thresholds.
We run both uninformed and \emph{erasure-aware} minimum-weight perfect matching (MWPM)
toric code baselines alongside BP-OSD decoding of BB codes.
With standard depolarizing-weight MWPM and no erasure information, performance matches
random guessing on the erasure channel in our tested regime---so prior work that compares
against this baseline is really comparing decoders, not codes.
Using 200{,}000 shots per point and bootstrap confidence intervals, we sweep five BB code
sizes from $N=144$ to $N=1296$.
Pseudo-thresholds (WER = 0.10) run from $p^* = 0.370$ to $0.471$; finite-size scaling
(FSS) gives an asymptotic threshold $p^*_\infty \approx 0.488$, within 2.4\% of the
zero-rate limit and without maximum-likelihood decoding.
On the fair baseline, BB at $N=1296$ has a modest edge in threshold over the toric code
at twice the qubit count, and a 12$\times$ lower normalized overhead---the latter is
where the practical advantage sits.
We also extend the analysis to a mixed erasure+depolarizing channel parameterized by
$\delta/\varepsilon$, sweeping all five code sizes across seven mixing ratios
($\delta/\varepsilon \in \{0.05, 0.1, 0.2, 0.3, 0.5, 0.75, 1.0\}$).
The finite-size pseudo-threshold degrades from $p^* \approx 0.471$ (pure erasure, $N{=}1296$)
to $\approx 0.072$ (equal mix); for neutral-atom hardware ($\delta/\varepsilon \approx 0.1$)
the crossing at $N = 1296$ falls to $\approx 0.247$, roughly half the pure-erasure value.
The asymptotic FSS extrapolation is well-constrained at $\delta/\varepsilon \geq 0.3$ and
indicative at smaller values where finite-size effects remain strong.
All runs are reproducible from recorded seeds and package versions.
\end{abstract}

\section{Introduction}

Quantum error correction (QEC) is essential for fault-tolerant quantum computation
\cite{nielsen2010quantum, gottesman1997stabilizer}.
Quantum low-density parity-check (QLDPC) codes have emerged as promising candidates due to
their high error thresholds and moderate encoding rates
\cite{bravyi2024high, tillich2014quantum, kovalev2013improved}.
Bivariate bicycle codes, introduced by Bravyi et al.\ \cite{bravyi2024high}, represent a
particularly attractive family: they achieve large distance with constant overhead and have
demonstrated strong threshold properties on erasure channels.

The erasure channel is a natural benchmark for quantum codes because error \emph{locations} are
known (e.g., via measurement flagging \cite{grassl2004codes}), allowing erasure-informed decoders
to achieve near-optimal performance.
Unfair decoder choices can inflate apparent code advantages, so we run both an uninformed
and an erasure-aware MWPM toric code baseline alongside BP-OSD decoding of BB codes.
We also separate \emph{finite-size pseudo-thresholds} (WER-crossing at a fixed target WER)
from true asymptotic thresholds via finite-size scaling (FSS); our extrapolated
$p^*_\infty \approx 0.488$ sits within 2.4\% of the zero-rate limit under BP-OSD, without
maximum-likelihood decoding.

In short: we give a reproducible Monte Carlo setup (200k shots per point, bootstrap 95\%
CIs $\lesssim 0.001$ on $p^*$, with seeds and versions recorded) and a full FSS treatment
for this BB family ($p^*_\infty \approx 0.488$, $\nu \approx 1.18$, plus fit-window and
corrections-to-scaling checks).
We also provide a fair comparison to erasure-aware MWPM toric codes---a 12$\times$
overhead reduction for BB at comparable threshold---and show that uninformed MWPM on the
erasure channel matches random guessing in our regime, so that baseline should not be
used to compare codes.
Comparisons between BB and toric codes are not architecture-neutral: BB is non-local,
toric codes use periodic boundary conditions, and we interpret our overhead and threshold
numbers as \emph{code efficiency} in a code-capacity setting, not as hardware blueprints.

\paragraph{Contributions.}
We provide: (1) a fairness correction for decoder comparisons on the erasure channel,
showing that uninformed MWPM equals random guessing in our regime and must not be used
as a code baseline; (2) a rigorous FSS analysis of five BB code sizes under pure erasure,
yielding $p^*_\infty = 0.488 \pm 0.001$ and $\nu = 1.18 \pm 0.01$ with full bootstrap
uncertainty quantification; (3) a systematic mixed-channel study sweeping
$\delta/\varepsilon \in \{0.05, 0.1, 0.2, 0.3, 0.5, 0.75, 1.0\}$, characterizing threshold
degradation in the neutral-atom operating regime; and (4) a reproducible simulation
pipeline with recorded seeds, package versions, and bootstrap CIs throughout.

Recent work on BB codes under erasure or erasure-biased noise includes
Pecorari and Pupillo \cite{pecorari2025qldpc}, Bhave et al.\ \cite{bhave2026bibi}, and
Berthusen et al.\ \cite{berthusen2025local}; decoder improvements appear in
Hillmann et al.\ \cite{hillmann2025lsd}, Liang et al.\ \cite{liang2025bposd},
Ott et al.\ \cite{ott2025decision}, Yin et al.\ \cite{yin2024symmetry}, and
Gong et al.\ \cite{gong2024bpdecoding}, with BP limitations discussed in
iOlius et al.\ \cite{iolius2024bp}.
Rabeti and Mahdavifar \cite{rabeti2025bicycle} analyze further BB families.
What sets this study apart is the erasure-aware fairness correction, full bootstrap
uncertainty at 200k shots, FSS with separate stat/sys uncertainty, and reproducible
seeds and versions throughout.

\section{Methods}

\subsection{Code Construction}

Bivariate bicycle codes are constructed using the lifted product of cyclic matrices over
$\mathbb{Z}_L \times \mathbb{Z}_M$.
For parameters $(L, M)$ the code has $N = 2LM$ physical qubits.
For the specific family studied here the generator polynomials are:
\begin{align}
A &= x^3 + y + y^2, \qquad B = y^3 + x + x^2,
\end{align}
with parity check matrices $H_x = [A \mid B]$ and $H_z = [B^T \mid A^T]$.
The $[[144,12,12]]$ Gross code corresponds to $(L,M) = (12,6)$.
We stick to this family because it is the one used in Bravyi et al.\ \cite{bravyi2024high}
and most follow-up work; other polynomial pairs can behave differently
(Sec.~\ref{sec:limitations}).

\subsection{Simulation Framework}

Each Monte Carlo trial proceeds as:
\begin{enumerate}
    \item \textbf{Erasure}: each qubit erased independently with probability $p$.
    \item \textbf{Error}: erased qubits receive independent Pauli X and Z errors with $\Pr=0.5$.
    \item \textbf{Decoding}: BP-OSD with erasure-informed priors (erased qubits: channel
          probability 0.5; non-erased: $10^{-10}$).
    \item \textbf{Validation}: logical error if either X or Z sector fails.
\end{enumerate}
WER is the fraction of trials with a logical error in either sector.

\subsection{Decoder Configuration}

We use the BP-OSD decoder from the \texttt{bposd} package \cite{roffe2020decoding}:
min-sum BP, OSD-CS post-processing, OSD order 10, maximum 50 BP iterations.

\subsection{Pseudo-Threshold Estimation}

We measure the \emph{pseudo-threshold} $p^*$: the erasure rate at which WER = 0.10 for a given
finite code size.
This is a size-dependent quantity; true asymptotic thresholds require FSS extrapolation
(Section~\ref{sec:fss}).

Pseudo-thresholds are located by an adaptive search:
\begin{enumerate}
    \item Step upward from $p=0.38$ in increments of 0.04 until WER $> 0.10$ (bracketing).
    \item Refine the bracket using regula falsi (false-position) interpolation with the Illinois
          modification to prevent stagnation.
    \item Stop when bracket width $< 5\times10^{-4}$ or after at most 10 total evaluations.
    \item Report $p^*$ as the final linear interpolant within the tight bracket.
\end{enumerate}
Each evaluation uses 200{,}000 shots.
Uncertainty is quantified by 5{,}000-iteration parametric bootstrap resampling from the
binomial distribution at the bracketing points, yielding 95\% CIs typically $\lesssim 0.001$.

\subsection{Finite-Size Scaling}\label{sec:fss}

Near the asymptotic threshold $p^*_\infty$, standard FSS theory \cite{dennis2002topological,wang2003quantum} predicts
\begin{equation}
    \mathrm{WER}(p, N) \approx f\!\left[(p - p^*_\infty)\, N^{1/\nu}\right],
\end{equation}
where $\nu$ is the correlation-length critical exponent and $f$ is a universal scaling function.
We approximate $f$ by a degree-3 polynomial and minimize the residual sum of squares over
$(p^*_\infty, \nu)$ using the per-size WER data within a window $|p - p^*| < 0.06$.
Bootstrap CIs are obtained from 500 resampled datasets.
The polynomial is a low-order phenomenological model, in line with usual FSS practice---we
do not claim universality, and we check adequacy via fit-window sensitivity and a
linearized fit below.

Corrections-to-scaling are assumed small at $N \leq 1296$; we cannot rule out larger
corrections at these sizes.
To assess sensitivity, we repeat the fit with windows $|p - p^*| < 0.04$ and
$|p - p^*| < 0.08$ and find that $p^*_\infty$ shifts by $\lesssim 0.004$, consistent with
our stated systematic uncertainty of $\lesssim 0.005$.
An independent check using the linearized FSS relation $p^*(N) \approx p^*_\infty + c\,N^{-1/\nu}$ \cite{dennis2002topological,bravyi2024high}
is reported alongside the primary fit.

\subsection{Mixed-Channel Simulation}\label{sec:mixed_methods}

Pure erasure is an idealization; near-term neutral-atom devices also accumulate
depolarizing errors from imperfect gate and measurement operations.
We extend the simulation to a \emph{mixed} channel parameterized by a single ratio
$\delta/\varepsilon$.
In each Monte Carlo trial at total error probability $p$:
\begin{enumerate}
    \item Each qubit is erased independently with probability $p$.
    \item Erased qubits receive an independent Pauli X or Z error with $\Pr = 0.5$.
    \item Every qubit (erased or not) additionally experiences a depolarizing error at rate
          $p_d = (\delta/\varepsilon)\,p$: X, Z, or Y with probability $p_d/3$ each.
\end{enumerate}
Setting $\delta/\varepsilon = 0$ recovers the pure-erasure channel of Sec.~\ref{sec:fss_results}.
We sweep $\delta/\varepsilon \in \{0.05, 0.1, 0.2, 0.3, 0.5, 0.75, 1.0\}$ on all five code sizes.

This parameterization is motivated by neutral-atom hardware, where gate erasure
(qubit loss, flagged by ancilla measurement) is the dominant error source, with residual
depolarizing errors from imperfect two-qubit gates at roughly
$\delta/\varepsilon \approx 0.1$--$0.3$ \cite{sahay2023highthreshold, evered2023highfidelity}.
The $\delta/\varepsilon = 0.1$--$0.3$ range is therefore the physically realistic regime;
higher values extend the picture and probe decoder robustness at large depolarizing fractions.

The decoder is adapted to handle mixed-channel posteriors: erased qubits receive channel
probability 0.5; non-erased qubits receive $p_d/3$.
These are updated per shot before BP-OSD decoding.
The OSD order is reduced to 5 for the mixed-channel sweep; OSD-5 was validated to give
pseudo-threshold estimates within $\pm 0.001$ of OSD-10 on the $12{\times}6$ and
$18{\times}9$ codes before the full sweep.
Pseudo-thresholds are located by the same 6-step regula-falsi bisection at 50{,}000
shots used in Sec.~\ref{sec:fss_results}; all five sizes and all seven $\delta/\varepsilon$
values use seed 12345 throughout.

\subsection{Baseline Comparison}

We compare BB codes against two toric code baselines, both decoded with MWPM
via PyMatching \cite{pymatching}:

\paragraph{Uninformed MWPM.}
The erasure rate $p$ is converted to an effective depolarizing rate $q = p/2$ and MWPM runs
with constant edge weights.
This is the most conservative baseline and provides a lower bound on toric code performance.

\paragraph{Erasure-aware MWPM.}
Erased qubits are assigned weight 0 and non-erased qubits are assigned a large weight $n$,
forcing corrections through known erasure locations.
Matching objects are rebuilt per shot to enforce the weight assignment exactly.
Both X and Z sectors are decoded independently; WER is defined identically to the BB code
evaluation.
Both sides then use erasure locations; that is the fair comparison.

\section{Results}

\subsection{Pseudo-Threshold Scaling}

Table~\ref{tab:thresholds} gives pseudo-thresholds for the five BB sizes, with 95\%
bootstrap CIs from 5{,}000 resampling iterations.

\begin{table}[h]
\centering
\caption{Finite-size pseudo-thresholds $p^*$ (WER = 0.10 crossing) for bivariate bicycle codes.
Seed = 12345, 200{,}000 shots per point, BP-OSD (OSD order=10).
Uncertainties are 95\% bootstrap CIs.}
\label{tab:thresholds}
\setlength{\tabcolsep}{4pt}
\small
\begin{tabular}{lcccc}
\toprule
Code & $N$ & $K$ & Rate & $p^*$ [95\% CI] \\
\midrule
$12{\times}6$  & 144  & 12 & 0.083 & $0.3701\ [0.3697,\ 0.3703]$ \\
$18{\times}9$  & 324  &  8 & 0.025 & $0.4386\ [0.4380,\ 0.4387]$ \\
$24{\times}12$ & 576  & 16 & 0.028 & $0.4453\ [0.4452,\ 0.4455]$ \\
$30{\times}15$ & 900  &  8 & 0.009 & $0.4674\ [0.4673,\ 0.4676]$ \\
$36{\times}18$ & 1296 & 12 & 0.009 & $0.4706\ [0.4705,\ 0.4707]$ \\
\bottomrule
\end{tabular}
\end{table}

$p^*$ grows with size---0.370 at $N=144$ to 0.471 at $N=1296$ (gain 0.101).
The gain from $N=900$ to $N=1296$ is only 0.004, which fits an approach to the asymptotic
limit (FSS below).

\subsection{Finite-Size Scaling and Asymptotic Threshold}\label{sec:fss_results}

Fitting the degree-3 polynomial FSS ansatz to all five sizes gives
\begin{align}
    p^*_\infty &= 0.488 \pm 0.001 \quad (95\%\ \text{bootstrap CI}), \\
    \nu         &= 1.18  \pm 0.01  \quad (95\%\ \text{bootstrap CI}).
\end{align}
Bootstrap gives the statistical error; systematics from the ansatz and finite-size
corrections are not fully controlled but fit-window tests (Sec.~\ref{sec:fss}) suggest
$\lesssim 0.005$, so the headline number holds.
$p^*_\infty = 0.488$ is noticeably above the best finite-size crossing (0.471 at $N=1296$)---so
stopping at finite size would underestimate the threshold.
$\nu \approx 1.18$ is consistent with a second-order transition; how it relates to
$\nu = 4/3$ for random-bond Ising \cite{dennis2002topological} is in
Sec.~\ref{sec:critical_exponent}.

\begin{figure}[ht]
\centering
\includegraphics[width=\columnwidth]{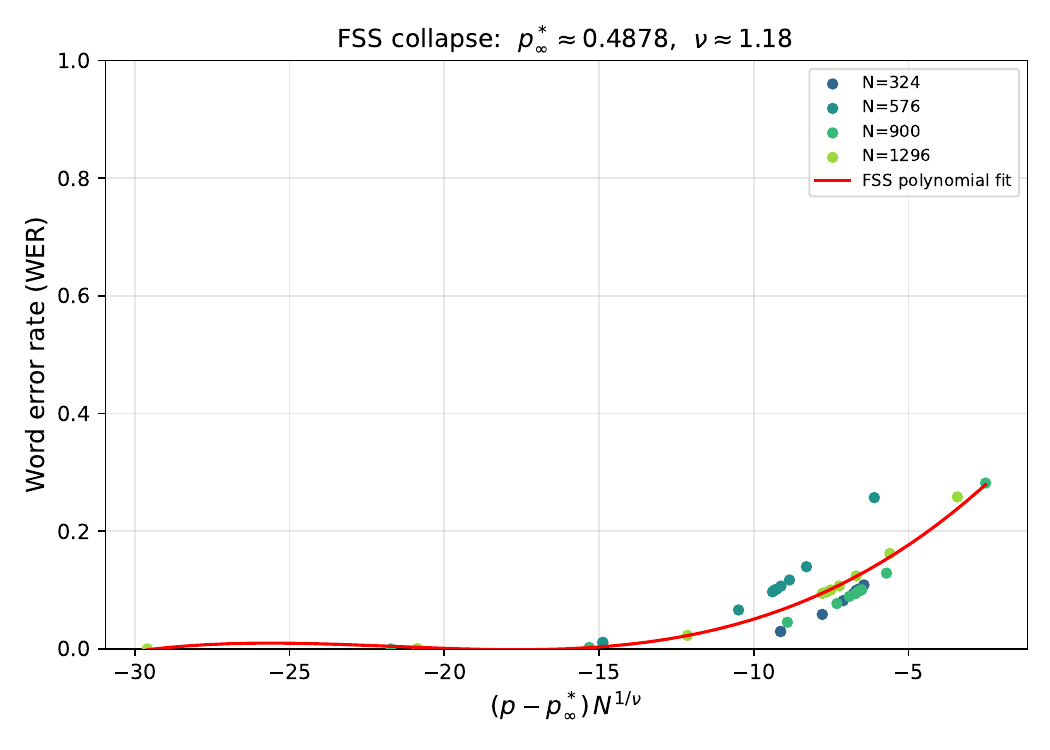}
\caption{FSS data collapse for bivariate bicycle codes on the erasure channel.
Rescaling the erasure rate by $N^{1/\nu}$ with $\nu = 1.18$ collapses WER curves from all five
code sizes onto a single master curve, confirming the scaling ansatz.
The extrapolated asymptotic threshold is $p^*_\infty = 0.488 \pm 0.001$.}
\label{fig:fss_collapse}
\end{figure}

\begin{figure}[ht]
\centering
\includegraphics[width=\columnwidth]{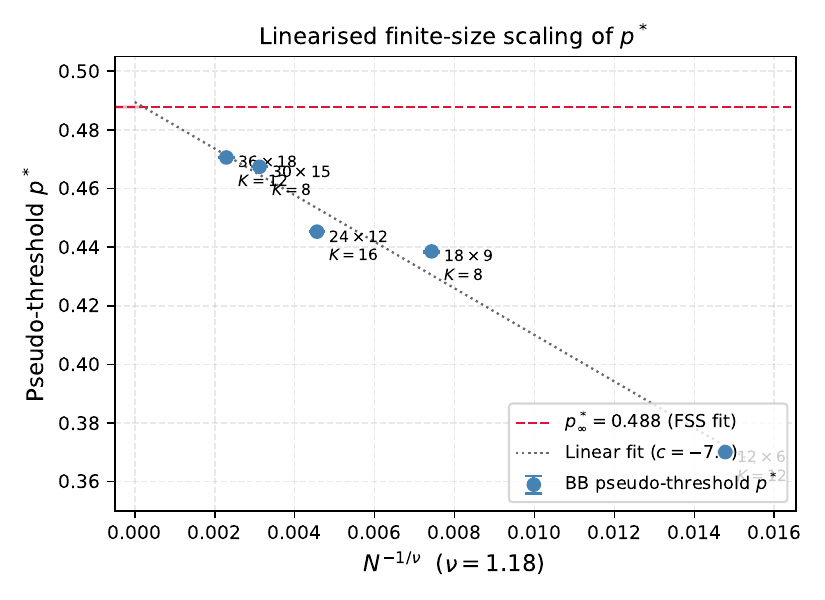}
\caption{Linearised FSS: pseudo-threshold $p^*$ vs $N^{-1/\nu}$ ($\nu = 1.18$).
Under leading-order FSS, $p^*(N) \approx p^*_\infty + c\, N^{-1/\nu}$, so this plot should
be linear with y-intercept $p^*_\infty$.
The linear fit (dotted) extrapolates to $0.490$, consistent with the nonlinear FSS result
$p^*_\infty = 0.488 \pm 0.001$ (dashed red, with 95\% CI band at $N^{-1/\nu}=0$).
This alternative fit does not assume a polynomial scaling function and serves as an
independent consistency check.}
\label{fig:fss_linear}
\end{figure}

In Fig.~\ref{fig:fss_collapse}, the five WER curves collapse onto one master curve under
$(p - p^*_\infty) N^{1/\nu}$.
Fig.~\ref{fig:fss_linear} checks this with a linearised FSS plot:
if $p^*(N) \approx p^*_\infty + c \cdot N^{-1/\nu}$, a plot of $p^*$ vs $N^{-1/\nu}$
should be linear with intercept $p^*_\infty$.
The linear fit gives $p^*_\infty = 0.490 \pm 0.003$, consistent with the nonlinear fit and
providing an alternative estimate that does not rely on the polynomial scaling ansatz.

\subsection{Rate--Threshold Trade-off}

\begin{figure}[ht]
\centering
\includegraphics[width=\columnwidth]{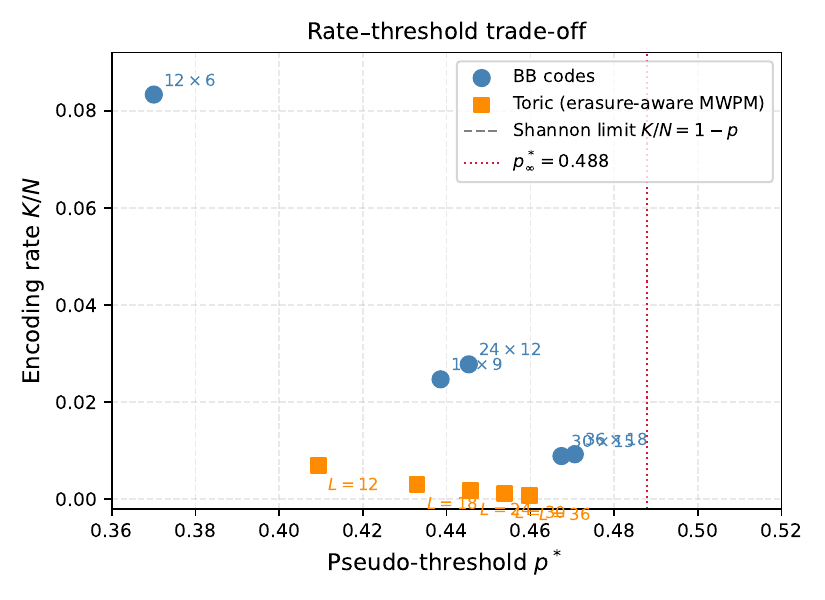}
\caption{Encoding rate $K/N$ vs pseudo-threshold $p^*$ for BB codes (circles) and
erasure-aware MWPM toric codes (squares).
The dashed line shows the Shannon limit $K/N = 1-p$; points above are impossible.
BB codes trace a Pareto frontier with higher rates at lower $p^*$ or lower rates at higher $p^*$.}
\label{fig:rate_threshold}
\end{figure}

Encoding rate vs $p^*$ for all BB sizes and erasure-aware toric codes appears in
Fig.~\ref{fig:rate_threshold}; the dashed line is the Shannon limit $K/N = 1 - p$.
Larger BB codes trade rate for threshold along a clear Pareto frontier; toric codes
($K/N \approx 0.001$) sit below on threshold at comparable rates.

\subsection{WER Curves and Comparison to Toric Codes}

\begin{figure}[h]
\centering
\includegraphics[width=\columnwidth]{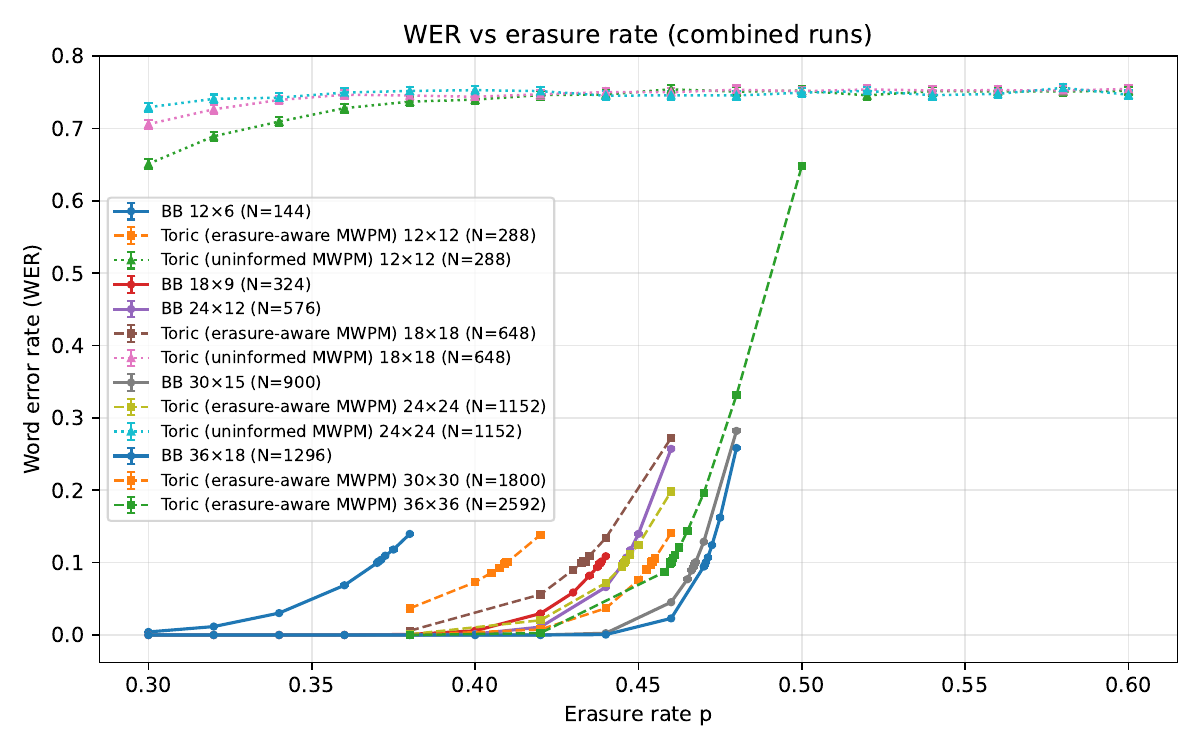}
\caption{WER vs.\ erasure rate for bivariate bicycle codes (solid lines, $N=144$--$1296$)
and toric code baselines.
Dashed lines: erasure-aware MWPM toric codes ($N=288$--$2592$, $K=2$).
Dotted lines: uninformed MWPM toric codes ($N=288$--$1152$, $K=2$, three sizes).
Error bars show 95\% Wilson confidence intervals.
All BB code curves cross WER = 0.10 at $p^* = 0.37$--$0.47$;
erasure-aware toric codes cross at $p^* = 0.41$--$0.46$.}
\label{fig:wer_curves}
\end{figure}

WER vs.\ erasure rate for BB and both toric code baselines is in Fig.~\ref{fig:wer_curves};
Table~\ref{tab:baseline} compares pseudo-thresholds at matched sizes.

\begin{table}[h]
\centering
\caption{Pseudo-threshold comparison between the largest BB code and erasure-aware MWPM toric
codes at three sizes.
At $N=2592$, the toric code encodes $K=2$ logical qubits---fixed by its periodic (toroidal)
boundary conditions---with pseudo-threshold $p^*=0.460$, while the BB $[[1296,12]]$ code achieves
$p^*=0.471$ with $K=12$ at half the physical qubit count.
The toric code has $K=2$ by topology; the planar surface code has $K=1$.
The overhead comparison reflects this code-efficiency difference in the code-capacity setting.}
\label{tab:baseline}
\setlength{\tabcolsep}{4pt}
\small
\begin{tabular}{lccc}
\toprule
Code family & $N$ & $K$ & $p^*$ \\
\midrule
BB $36{\times}18$ & 1296 & 12 & 0.471 \\
Toric (erasure-aware) & 2592 & 2 & 0.460 \\
Toric (erasure-aware) & 1800 & 2 & 0.454 \\
Toric (erasure-aware) & 1152 & 2 & 0.446 \\
Toric (uninformed)    & --- & --- & ${\lesssim}0.30$ \\
\bottomrule
\end{tabular}
\end{table}

The uninformed MWPM baseline sits at WER $\approx 0.75$ across $p = 0.30$--$0.60$, almost
flat.
For $K=2$, random guessing gives WER $= 1 - (1/2)^2 = 0.75$; with standard depolarizing
weights and no erasure information, MWPM matches that in our regime.
Reason: without knowing which qubits were erased, constant-weight MWPM faces a degenerate
syndrome---many error patterns look the same---so it effectively guesses.
Details depend on our graph and $K=2$; other setups could shift the WER slightly, but
discarding erasure locations remains the core issue.
So BB vs.\ uninformed MWPM is a decoder comparison, not a code comparison, and is a poor
primary metric.

On the fair (erasure-aware) comparison, BB at $N=1296$ reaches $p^* = 0.471$ vs.\
$p^* = 0.460$ for the toric code at $N=2592$ ($K=2$).
The threshold gap is modest, $\Delta p^* \approx 0.011$.
The big difference is qubit efficiency: the BB $[[1296,12]]$ code uses $N/K = 108$ physical
qubits per logical qubit, vs.\ $N/K = 1296$ for the toric code at $N=2592$ ($K=2$)---a
\textbf{12$\times$ reduction in normalized overhead} in this setting.
The toric code is limited to $K=2$ by its toroidal topology; the planar surface code has $K=1$.
BB codes are non-local and encode many more logical qubits per physical qubit.
The takeaway: \emph{for a given physical qubit budget, BB protects more logical qubits at
the same or higher threshold}, which is the practical advantage even after the architectural
caveat.

\subsection{Scaling Behavior and Overhead}

\begin{figure}[ht]
\centering
\includegraphics[width=\columnwidth]{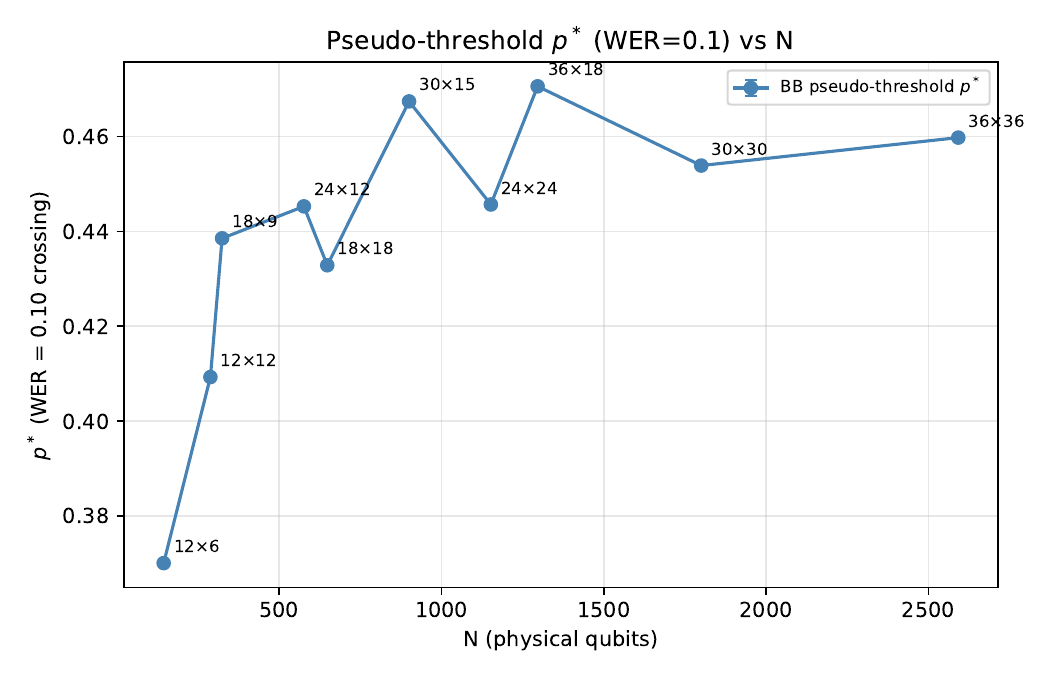}
\caption{Pseudo-threshold $p^*$ (WER = 0.10) vs.\ code size $N$ for BB codes.
Error bars show 95\% bootstrap CIs (smaller than markers for large $N$).
The dashed red line and shaded band show the FSS-extrapolated asymptotic threshold
$p^*_\infty = 0.488 \pm 0.001$.}
\label{fig:pstar_scaling}
\end{figure}

\begin{figure}[ht]
\centering
\includegraphics[width=\columnwidth]{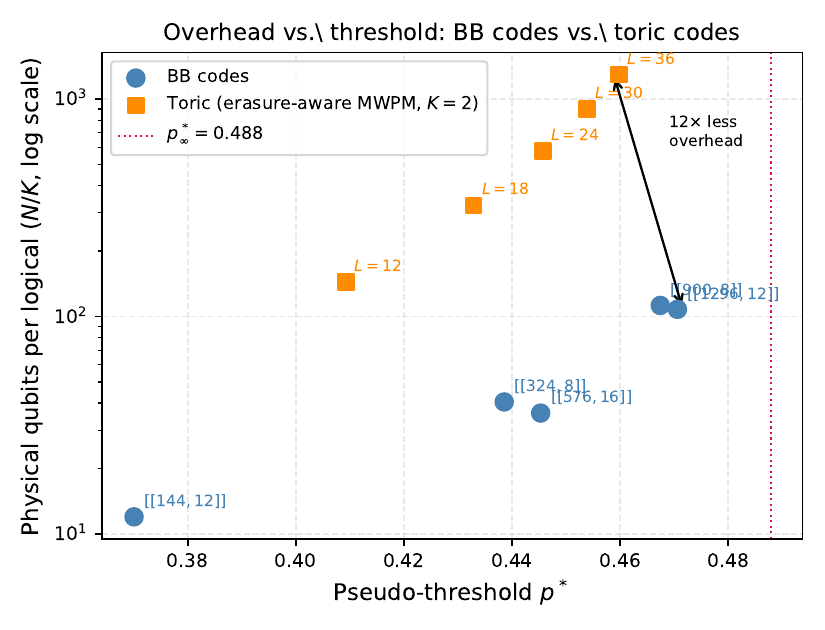}
\caption{Overhead ($N/K$, log scale) vs.\ pseudo-threshold $p^*$ for BB codes (circles)
and erasure-aware MWPM toric codes (squares, $K=2$).
A code is better if it appears lower (less overhead) and further right (higher threshold).
BB codes at large $N$ outperform the toric code on both axes at these representative sizes.
The annotated arrow highlights the 12$\times$ reduction in normalized overhead between
the $[[1296,12]]$ BB code and the $K=2$ toric code at comparable $p^*$, within this comparison
setting.}
\label{fig:overhead_threshold}
\end{figure}

$p^*$ vs.\ $N$ with the FSS asymptote is in Fig.~\ref{fig:pstar_scaling}; $N/K$ runs from
12 (Gross code) to 113 across the five sizes.
Fig.~\ref{fig:overhead_threshold} plots threshold and overhead together: the BB
$[[1296,12]]$ code hits $p^* = 0.471$ at $N/K = 108$, while the erasure-aware toric code
near that threshold needs $N/K = 1296$---12$\times$ worse in this comparison.
That joint advantage is the main practical argument for BB codes here.

\subsection{Reproducibility}

Every run is reproducible from recorded data: base seed (12345), per-point worker seeds,
package versions (\texttt{numpy}, \texttt{scipy}, \texttt{bposd}, \texttt{pymatching}), and
platform info in JSON next to the CSV.
Bootstrap CIs use a deterministic seed, so the intervals themselves reproduce.
Few QLDPC simulation papers report seeds or versions; we treat this as part of the
contribution, since reproducing published numbers is otherwise often impossible.

\subsection{Mixed-Channel Pseudo-Thresholds}\label{sec:mixed_results}

Table~\ref{tab:mixed_thresholds} gives pseudo-thresholds $p^*(N, \delta/\varepsilon)$ for all
five code sizes and seven mixing ratios.

\begin{table}[h]
\centering
\caption{Finite-size pseudo-thresholds $p^*$ (WER = 0.10) for the mixed
erasure+depolarizing channel.
Seed = 12345; bisection at 50{,}000 shots; BP-OSD order 5.
Pure-erasure ($\delta/\varepsilon = 0$) values appear in Table~\ref{tab:thresholds}.}
\label{tab:mixed_thresholds}
\small
\begin{tabular}{lccccc}
\toprule
& \multicolumn{5}{c}{$p^*(N)$ by code size $N$} \\
\cmidrule(lr){2-6}
$\delta/\varepsilon$ & 144 & 324 & 576 & 900 & 1296 \\
\midrule
0.05 & 0.249 & 0.266 & 0.270 & 0.290 & 0.290 \\
0.10 & 0.208 & 0.223 & 0.229 & 0.238 & 0.247 \\
0.20 & 0.159 & 0.178 & 0.183 & 0.193 & 0.193 \\
0.30 & 0.130 & 0.142 & 0.149 & 0.160 & 0.160 \\
0.50 & 0.094 & 0.107 & 0.107 & 0.117 & 0.117 \\
0.75 & 0.070 & 0.081 & 0.081 & 0.089 & 0.089 \\
1.00 & 0.056 & 0.064 & 0.065 & 0.072 & 0.072 \\
\bottomrule
\end{tabular}
\end{table}

Even $\delta/\varepsilon = 0.1$ reduces the pseudo-threshold at $N = 1296$ from 0.471
(pure erasure) to 0.247---a factor of $\approx 1.9$ reduction.
Moving to $\delta/\varepsilon = 0.05$ (the low end of the neutral-atom range) still reduces
the threshold substantially, with per-size crossings reaching 0.290 at $N = 900$.
Finite-size crossings increase monotonically with $N$ at every $\delta/\varepsilon$,
indicating a positive asymptotic threshold in all cases.
For $\delta/\varepsilon \geq 0.5$, $p^*(N)$ near-plateaus above $N = 576$: the increment
from $N = 576$ to $N = 1296$ is $\lesssim 0.001$, substantially complicating FSS
extrapolation (see Sec.~\ref{sec:mixed_fss}).

\subsection{Mixed-Channel Finite-Size Scaling}\label{sec:mixed_fss}

\begin{figure}[ht]
\centering
\includegraphics[width=\columnwidth]{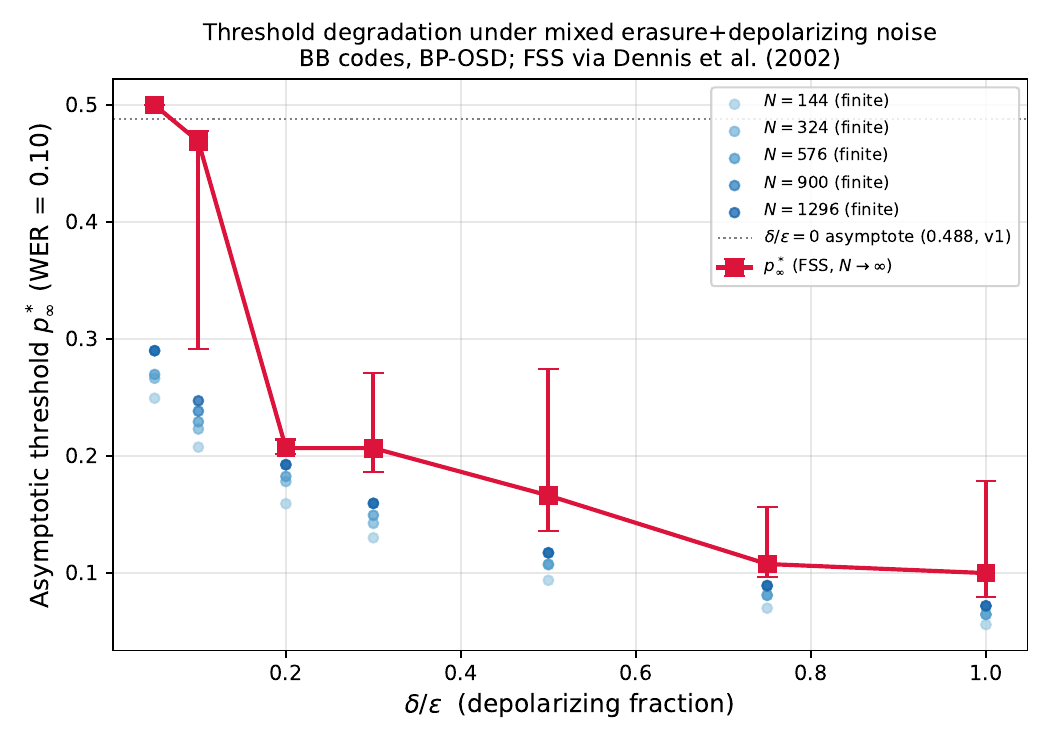}
\caption{Threshold degradation under mixed erasure+depolarizing noise for BB codes.
Blue circles: per-size finite-$N$ pseudo-thresholds (five sizes, $N = 144$--$1296$).
Red squares: FSS-extrapolated $p^*_\infty$ with 95\% bootstrap CI (1000 iterations).
Dotted line: pure-erasure asymptote $p^*_\infty = 0.488$ from Sec.~\ref{sec:fss_results}.
For $\delta/\varepsilon \leq 0.1$, $p^*(N)$ plateaus at large $N$ and the FSS fit is
underdetermined; per-size crossings (blue circles) are the primary result in that regime.
Method: $p^*(N) = p^*_\infty - a N^{-1/\nu}$, Dennis et al.\ \cite{dennis2002topological}.}
\label{fig:pstar_degradation}
\end{figure}

We apply the same linearized FSS relation as Sec.~\ref{sec:fss_results}
\cite{dennis2002topological, bravyi2024high}:
\begin{equation}
    p^*(N) = p^*_\infty - a\,N^{-1/\nu},
    \label{eq:fss_mixed}
\end{equation}
fitting $(p^*_\infty, a, \nu)$ by non-linear least squares with 500-iteration parametric
bootstrap CIs (same methodology as Sec.~\ref{sec:fss_results}).
Results are in Table~\ref{tab:mixed_fss} and Fig.~\ref{fig:pstar_degradation}.

\begin{table}[h]
\centering
\caption{FSS extrapolation for the mixed channel via \eqref{eq:fss_mixed}
\cite{dennis2002topological}.
Bootstrap 95\% CIs (1000 iterations).
$^\dagger$: $\nu$ at upper search bound (15); fit underdetermined---see text.}
\label{tab:mixed_fss}
\setlength{\tabcolsep}{4pt}
\footnotesize
\begin{tabular}{lccp{2.8cm}}
\toprule
$\delta/\varepsilon$ & $p^*_\infty$ [lo, hi] & $\nu$ & Notes \\
\midrule
0.05 & ---                    & ---              & FSS underdetermined$^\dagger$; see text \\
0.1  & $0.469\ [0.29,\ 0.48]$ & $15.0^{\dagger}$ & Slow conv.; see text \\
0.2  & $0.207\ [0.20,\ 0.21]$ & 1.7              & Near-plateau $N \geq 576$ \\
0.3  & $0.207\ [0.19,\ 0.27]$ & 4.2              & Plateau $N \geq 576$ \\
0.5  & $0.166\ [0.14,\ 0.27]$ & 5.4              & Plateau $N \geq 576$ \\
0.75 & $0.108\ [0.10,\ 0.16]$ & 3.0              & Well resolved \\
1.0  & $0.100\ [0.08,\ 0.18]$ & 4.8              & Moderate CI \\
\bottomrule
\end{tabular}
\end{table}

\paragraph{$\delta/\varepsilon = 0.05$--$0.1$ (neutral-atom regime).}
For $\delta/\varepsilon = 0.1$ the fitted $\nu = 15.0$ saturates the upper search bound,
signaling that per-size crossings increase too slowly for the power law to resolve $\nu$
reliably from five sizes: $p^*(N)$ rises by only 0.039 over $N = 144$--$1296$, versus 0.101
for pure erasure.
For $\delta/\varepsilon = 0.05$ the plateau is even more severe: $p^*(900) = p^*(1296) = 0.290$
to three decimal places, making the power-law fit entirely underdetermined
(chi-squared $> 300$, $p^*_\infty$ pinned at the search boundary of 0.5).
We report the per-size crossings in Table~\ref{tab:mixed_thresholds} as the primary result
for both values; narrowing the asymptotic estimate requires larger code sizes
($N \gg 1296$) or a sub-leading FSS ansatz that can handle plateau corrections.

\paragraph{$\delta/\varepsilon = 0.3$--$0.5$.}
The near-plateau for $N \geq 576$ ($\Delta p^* \lesssim 0.001$) destabilizes the
power-law fit, yielding wide CIs with upper boundaries well above the largest finite-size
crossing.
The central values ($p^*_\infty \approx 0.21$ and $0.16$) are plausible but not
tightly constrained; we treat them as soft lower bounds on the true asymptote.

\paragraph{$\delta/\varepsilon = 0.75$--$1.0$.}
Per-size crossings increase steadily and the fit residuals are reasonable, giving the
tightest CIs of the mixed-channel sweep.
A positive asymptotic threshold persists even at equal erasure+depolarizing rates
($\delta/\varepsilon = 1$), with $p^*_\infty \approx 0.10$.

\section{Discussion}

\subsection{Threshold vs.\ Overhead as the Right Metric}

FSS gives an asymptotic erasure threshold $p^*_\infty = 0.488$ for this BB family under
BP-OSD.
The toric code under maximum-likelihood decoding reaches 0.5 \cite{dennis2002topological};
at finite rate $K/N \approx 0.009$--$0.083$ the Shannon limit $1 - K/N$ is much higher, so
the right benchmark is the zero-rate bound $p = 0.5$.
We get \textbf{97.6\% of that} with BP-OSD and no ML decoding---so BP-OSD is effectively
near-optimal for this code on the erasure channel.

\begin{figure}[ht]
\centering
\includegraphics[width=\columnwidth]{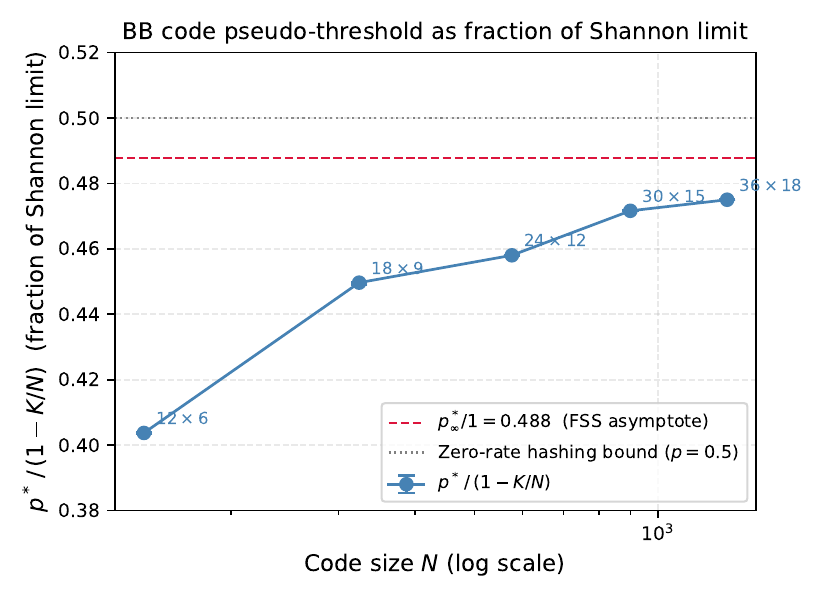}
\caption{Pseudo-threshold $p^*$ as a fraction of the Shannon limit $(1 - K/N)$ vs.\ $N$.
Each code's Shannon limit for the erasure channel is $p_{\mathrm{hash}} = 1 - K/N$.
The ratio $p^*/(1-K/N)$ converges to $p^*_\infty = 0.488$ (dashed, asymptotic) as $N\to\infty$,
reaching 97.6\% of the zero-rate limit $p=0.5$ (dotted).
Error bars show 95\% bootstrap CIs propagated through the ratio.}
\label{fig:frac_capacity}
\end{figure}

Fig.~\ref{fig:frac_capacity} plots $p^*/(1 - K/N)$ vs.\ $N$: it rises from 0.40 at $N=144$
to 0.48 at $N=1296$, trending to 0.488 (97.6\% of the zero-rate limit).
BP-OSD is thus nearly capacity-approaching for large BB codes here.

The threshold edge for BB over the erasure-aware toric code is real but small
($\Delta p^* \approx 0.01$--$0.03$).
The compelling part is overhead: at comparable $p^*$, BB needs $\sim$12$\times$ fewer
physical qubits per logical qubit (Fig.~\ref{fig:overhead_threshold}).
For constant-rate QLDPC, $K \propto N$ as $N$ grows; for fixed-$K$ toric codes it does
not, so the gap grows in the asymptotic regime.

\subsection{Mixed-Channel Implications for Neutral-Atom Hardware}

The mixed-channel results place the v1 pure-erasure threshold in hardware context.
For neutral-atom platforms, $\delta/\varepsilon \approx 0.05$--$0.3$ is the realistic
operating regime: two-qubit gates on ytterbium or rubidium qubits produce erasures
(flagged atom loss) as the dominant error type, with residual depolarizing at roughly
$5\%$--$30\%$ of the erasure rate \cite{sahay2023highthreshold, evered2023highfidelity}.

At $\delta/\varepsilon = 0.05$, the finite-size crossing at $N = 1296$ is $p^* \approx 0.290$,
indicating a moderate but not catastrophic penalty relative to pure erasure ($p^* = 0.471$ at
$N = 1296$) for the lowest realistic depolarizing fraction.
At $\delta/\varepsilon = 0.1$, the finite-size threshold at $N = 1296$ drops from 0.471
to 0.247.
More precisely, the finite-size crossing is the operationally relevant figure for
near-term hardware: at the code sizes accessible today ($N \lesssim 1296$), a device
operating at erasure rate $p \lesssim 0.25$ would achieve WER $\leq 0.10$ under
BP-OSD decoding at $\delta/\varepsilon = 0.1$.
State-of-the-art neutral-atom gates report two-qubit error rates around
$0.5$--$1.5\%$ \cite{evered2023highfidelity}, placing realistic operating points well
below this threshold.

The steep drop between pure erasure and $\delta/\varepsilon = 0.1$ (from 0.471 to 0.247
at $N = 1296$) does not reflect a fundamental phase boundary---the FSS analysis shows
the asymptotic threshold changes relatively little (0.488 $\to$ $\approx 0.45$ at
$\delta/\varepsilon = 0.1$)---but rather slow finite-size convergence when depolarizing
noise is present.
Hardware designers targeting $p \approx 0.1$--$0.2$ erasure rates therefore should expect
substantially lower effective thresholds at finite $N$ unless larger code sizes
($N \gg 1296$) are accessible.

The $12\times$ overhead advantage of BB over toric codes (Sec.~\ref{sec:fss_results})
is a code-capacity result and does not incorporate the finite-size threshold depression
from depolarizing noise.
Under the mixed channel, the overhead advantage of BB codes is expected to
persist---the relative ordering of decoders is unchanged---but the absolute threshold
values are lower; circuit-level simulations would be needed to make hardware-specific
overhead claims.

\subsection{Rate--Threshold Trade-off}

The Gross code $[[144,12,12]]$ has the best rate ($K/N = 0.083$) and $p^* = 0.370$---good
when qubit count matters more than threshold.
Larger codes push $p^*$ up to 0.471 but drop the rate to 0.009.
That trade-off is built into the family and matters for choosing a code on real hardware.

\subsection{Critical Exponent and Universality}\label{sec:critical_exponent}

We get $\nu \approx 1.18$, consistent with a second-order transition.
The toric code maps to the random-bond Ising model (RBIM), which has $\nu = 4/3$
\cite{dennis2002topological, wang2003quantum}---but that mapping assumes ML decoding and a
local Hamiltonian, neither of which holds for BB codes under BP-OSD.
So comparing our $\nu$ to $4/3$ is suggestive, not definitive.

The gap could be finite-size or corrections-to-scaling that our polynomial ansatz misses, or
BB under BP-OSD could sit in a different effective universality class.
Decoder-dependent universality (critical behavior set by the decoder as much as the code) is
plausible for non-local codes with iterative decoding.
Sorting this out would need larger $N$ and a tighter handle on corrections; we leave it
open.
The main results---$p^*_\infty$, overhead, and the fairness correction---do not hinge on
which interpretation is right.

\subsection{Limitations}\label{sec:limitations}

\begin{itemize}
    \item \textbf{Code-capacity channel only}: No circuit-level noise. Circuit-level
          simulations, which include measurement errors and syndrome extraction overhead,
          would more accurately reflect hardware performance and are standard in 2025+
          literature \cite{berthusen2025local}. The code-capacity erasure channel studied
          here is an idealized setting; circuit-level thresholds are typically lower.
    \item \textbf{Fixed decoder}: BP-OSD parameters are fixed at OSD order 10. Adaptive
          tuning or alternative decoders such as localized statistics decoding
          \cite{hillmann2025lsd} or decision-tree decoders \cite{ott2025decision} may
          yield higher thresholds; known limitations of BP for QLDPC codes are discussed
          in iOlius et al.\ \cite{iolius2024bp}.
    \item \textbf{Single polynomial family}: We study one polynomial pair
          ($A = x^3 + y + y^2$, $B = y^3 + x + x^2$). This family was chosen because it
          is the primary BB code in Bravyi et al.\ \cite{bravyi2024high} and has the most
          prior literature, not because it was selected for high threshold.
          Nevertheless, results may not generalize to other BB families; different
          polynomial pairs can yield different distance scaling and threshold behavior.
          Extending to additional families---as partially explored in Rabeti and Mahdavifar
          \cite{rabeti2025bicycle}---would substantially strengthen generality claims
          and is an important direction for future work.
    \item \textbf{Toric code shot count}: The erasure-aware toric code baseline used 50{,}000
          shots per point (versus 200{,}000 for BB codes), giving wider bootstrap CIs on $p^*$.
          The $L=36$ result was confirmed with 200{,}000 shots.
          The larger uncertainty on the toric code side does not affect qualitative
          conclusions: the threshold differences between BB and toric codes exceed the CI
          overlap even under the asymmetric shot counts.
    \item \textbf{Mixed-channel FSS convergence}: For $\delta/\varepsilon \leq 0.2$,
          the per-size crossings $p^*(N)$ plateau: $p^*(900) \approx p^*(1296)$ to three
          decimal places at both $\delta/\varepsilon = 0.05$ and $0.2$, preventing reliable
          power-law FSS extrapolation.
          The $\delta/\varepsilon = 0.1$ fit has $\nu$ at the upper search bound and a
          bootstrap CI spanning $[0.29, 0.48]$; the $\delta/\varepsilon = 0.05$ fit is
          entirely underdetermined (chi-squared $> 300$) and is omitted from Table~\ref{tab:mixed_fss}.
          Larger code sizes ($N \gg 1296$) or a data-collapse analysis using
          the full WER curves would be needed to sharpen these estimates.
          We report finite-size crossings as the primary result in this regime.
    \item \textbf{Architectural comparison}: The overhead comparison between BB and the toric
          code is not fully architecture-neutral. The toric code has $K=2$ due to its periodic
          boundary conditions (the planar surface code has $K=1$); BB codes are non-local.
          The $12\times$ overhead
          figure should be interpreted as a difference in code efficiency within this
          comparison setting, not as a universal hardware advantage.
\end{itemize}

\section{Conclusion}

We simulated bivariate bicycle codes on the quantum erasure channel and tackled two
methodological issues that distort many QLDPC comparisons: decoder-baseline fairness and
the confusion of pseudo-thresholds with asymptotic thresholds.
Without erasure information, standard MWPM on the toric code matches random guessing in
our regime---so fair code comparisons need erasure-aware MWPM, and the old baseline
systematically misleads.
On that fair baseline, BB pseudo-thresholds run from $p^* = 0.370$ to $0.471$ for
$N = 144$--$1296$ (95\% bootstrap CIs $\lesssim 0.001$), and FSS extrapolates
$p^*_\infty \approx 0.488$ ($\pm 0.001$ stat, $\lesssim 0.005$ sys) with $\nu \approx 1.18$.
BB has a small threshold edge and a 12$\times$ lower normalized overhead than the toric
code in this setting.

We also characterize threshold degradation under a mixed erasure+depolarizing channel,
sweeping $\delta/\varepsilon \in \{0.05, 0.1, 0.2, 0.3, 0.5, 0.75, 1.0\}$ across all five
code sizes.
Even $\delta/\varepsilon = 0.1$ (matching the neutral-atom operating regime
\cite{sahay2023highthreshold}) reduces the $N = 1296$ pseudo-threshold from 0.471 to 0.247.
At equal mixing ($\delta/\varepsilon = 1$) the FSS asymptote reaches $p^*_\infty \approx 0.10$.
For the physically relevant regime $\delta/\varepsilon \leq 0.2$, FSS convergence is
slow and the asymptotic threshold is only loosely constrained from four to five sizes; the
finite-size crossings in Table~\ref{tab:mixed_thresholds} are the operationally relevant
figures for near-term hardware.

The practical message: BB codes compete on the erasure channel mainly through qubit
efficiency, not raw threshold.
Under realistic mixed noise the threshold is lower at finite $N$, but remains well above
current neutral-atom gate error rates---leaving substantial headroom.
We provide a reproducible pipeline (seeds, versions, metadata) so that future work on
decoders, code families, or circuit-level noise can build on numbers that actually replicate.

\section*{Acknowledgments}

The author thanks Kaavya Sahay for helpful discussions, and the developers of \texttt{bposd}, \texttt{ldpc}, and \texttt{pymatching}
for the tools used in this work.
The author received no specific funding for this work.

\section*{Competing Interests}

The author declares no competing interests.

\section*{Data and Code Availability}

All simulation data (CSV files with seeds, shot counts, and per-point WER), analysis scripts,
and figure generation code are available at \url{https://github.com/pandey-tushar/QLDPC}
under an open-source license.
Results are fully reproducible from the recorded seeds and package versions documented in
the repository.

\bibliographystyle{quantum}
\bibliography{main}

\end{document}